\documentclass[apj]{emulateapj}
\usepackage{apjfonts}
\usepackage{graphicx}
\usepackage{color}
\usepackage{amsbsy}
\usepackage{mathpazo,bm}
\usepackage{amsmath}
\usepackage{multirow}
\usepackage{natbib}
\usepackage{amssymb}
\newlength{\hfwidth}
\newlength{\hfwidthsingle}
\newcommand{\pderiv}[2]{\frac{\partial #1}{\partial #2}}

\newcommand{\vt}[1]{\mathbf{#1}}       
\renewcommand{\v}[1]{{\boldsymbol{#1}}} 
\def\white#1{\textcolor{white}{#1}}

\definecolor{brown}{rgb}{0.42,0.24,0.07}

\newcommand{\del}{\v{\nabla}}
\newcommand{\grad}{\del}
\newcommand{\Div}{\del\cdot}

\newcommand{\Laplace}{\nabla^2}
\newcommand{\Eq}[1]{Eq. (\ref{#1})}
\newcommand{\Eqs}[2]{Eqs. (\ref{#1}) and~(\ref{#2})}
\newcommand{\Eqss}[2]{Eqs. (\ref{#1})--(\ref{#2})}
\newcommand{\eq}[1]{\Eq{#1}}
\newcommand{\eqs}[2]{\Eqs{#1}{#2}}
\newcommand{\eqss}[2]{\Eqss{#1}{#2}}
\newcommand{\eqp}[1]{(Eq. \ref{#1})}

\newcommand{\Fig}[1]{Fig.~\ref{#1}}
\newcommand{\fig}[1]{\Fig{#1}}

\newcommand{\sect}[1]{Sect.~\ref{#1}}

\newcommand{\beq}{\begin{equation}}
\newcommand{\eeq}{\end{equation}}
\newcommand{\beqn}{\begin{eqnarray}}
\newcommand{\eeqn}{\end{eqnarray}}
\newcommand{\epsp}{\xi_{_{+}}}
\newcommand{\epsm}{\xi_{_{-}}}
\newcommand{\tilchi}{\tilde\chi}
\newcommand{\tauf}{\tau_{\rm f}}

\newcommand{\St}{{\rm St}}

\shorttitle{Dust in disk vortices}
\shortauthors{Lyra \& Lin}

\slugcomment{Draft version}

\begin{document}

\title{Steady state of dust distributions in disk vortices:\\
Observational predictions and applications to transitional disks}
\author{Wladimir Lyra\altaffilmark{1,2,3,$\star$} and Min-Kai Lin\altaffilmark{4,$\star$}}
\email{wlyra@caltech.edu, \ mklin924@cita.utoronto.ca}
\altaffiltext{1}{Jet Propulsion Laboratory, California Institute of Technology, 4800 Oak Grove Drive, Pasadena, CA, 91109, USA}
\altaffiltext{2}{Division of Geological \& Planetary Sciences, California Institute of Technology, 1200 E California Blvd MC 150-21, Pasadena, CA 91125 USA}
\altaffiltext{3}{Sagan Fellow}
\altaffiltext{4}{Canadian Institute for Theoretical Astrophysics , 60
  St. George Street, Toronto, Ontario, M5S 3H8, Canada}
\altaffiltext{$\star$}{Both authors contributed equally to this work}

\begin{abstract}
The Atacama Large Millimeter Array (ALMA) has been returning images of transitional disks in which large asymmetries are seen in the distribution of 
mm-sized dust in the outer disk. The explanation in vogue borrows from the vortex literature by suggesting 
that these asymmetries are the result of dust trapping in giant vortices, excited via Rossby wave instability (RWI) 
at planetary gap edges. Due to the drag force, dust trapped in vortices will accumulate 
in the center, and diffusion is needed to maintain a steady state over the lifetime of the disk. While previous work 
derived semi-analytical models of the process, in this paper we
provide analytical steady-state solutions. Exact solutions exist for certain vortex models.
The solution is determined by the vortex rotation profile, the gas scale height, the 
vortex aspect ratio, and the ratio of dust diffusion to gas-dust friction. In principle, all these quantities can be derived
from observations, which would give validation of the model, also giving constrains on the strength of the turbulence 
inside the vortex core. Based on our solution, we derive quantities such as the gas-dust contrast, the trapped dust mass,
and the dust contrast at the same orbital location. We apply our model to the recently imaged Oph IRS 48 system, finding 
values within the range of the observational uncertainties.
\end{abstract}

\section{Introduction}
\label{sect:introduction}

Transitional disks are a class of circumstellar disks that lack a
significant near-infrared (1-5$\mu$m) excess, while showing steep
slopes in mid-infrared (5-20$\mu$m) and
far-infrared ($>$20$\mu$m) excesses typical of classical T-Tauri disks
\citep{Strom89,Skrutskie90,Gauvin-Strom92,Wolk-Walter96,Calvet02,Calvet05,Muzerolle06,Sicilia06,Currie09,Currie-Sicilia11}. 
This ``opacity hole''  implies absence of optically thick warm dust in the inner disk, with a dust
wall generating the mid-IR emission, followed by cold dust in the
outer disk.  This, together with their age (in the 1-10 Myr range, see
e.g. \citealt{Currie10} for a review) provide strong evidence that these are
objects caught in the evolutionary stage between gas-rich 
primordial and gas-poor debris disks, hence the name. 

Explanations for the opacity hole generally fall in four distinct
categories. These are, namely, grain growth and dust settling \citep{Brauer07,Dominik-Dullemond08,Zsom11,Birnstiel12}, photoevaporation
\citep{Alexander06,Cieza08,Pascucci-Sterzik09,Owen10}, 
dynamical interaction with close stellar or substellar companions
\citep{Ireland-Kraus08}, and planet
formation via dust locking \citep{Safronov69,Lyttleton72,Goldreich-Ward73,Youdin-Shu02,Johansen07} and gap
carving \citep{Papaloizou-Lin84,Lin-Papaloizou86a,Lin-Papaloizou86b,Bryden99,Paardekooper-Mellema04,Quillen04,Najita07,Andrews11}. 
Analyses of individual disks \citep{Calvet04,Calvet05,Espaillat08} tend to favor one process over another, and even census
studies of statistically significant samples of disks find one process
to be dominant \citep{Najita07,Cieza08}. These
seemingly conflicting results in fact illustrate the
heterogeneity of transitional disks, where a combination of all
suggested processes are needed to explain the rich diversity observed 
\citep{Cieza10,Muzerolle10,Merin10,Rosotti13,Clarke-Owen13}.   

Recently, high angular resolution imaging of the outer regions of transitional 
disks have become available, showing a myriad of puzzling asymmetries 
that beg for explanation. These come in the shape of spiral arms
\citep{Pietu05,Corder05,Muto12,Tang12},  elliptical dust walls
\citep{Isella12}, and non-axisymmetric dust clouds
\citep{Oppenheimer08,Brown09,Casassus12}. In particular, giant horseshoe-shaped dust distributions are seen in images obtained with 
the Combined Array for Research in Millimeter-wave Astronomy 
(CARMA, \citealt{Isella13}) and with the Atacama Large Millimeter Array (ALMA, \citealt{Casassus13,vanderMarel13}). The planet interpretation is particularly attractive for 
explaining these asymmetries, since they generally match the range of structures predicted by hydrodynamical 
models of planet-disk interaction. 

A deep gap is one of these expected structures, as the planet tides expel material from the vicinity of its orbit 
\citep{Papaloizou-Lin84,Lin-Papaloizou86a,Lin-Papaloizou86b,Nelson00,Masset-Snellgrove01,Paardekooper-Mellema04,Quillen04,deValBorro06,Klahr-Kley06,Lyra09a,Zhu11,Kley12,Kley-Nelson12}. The gas gap walls constitute steep pressure gradients, that, by
modifying the rotational profile locally, are prone to excite what has been called Rossby wave
instability \citep[RWI,][]{Lovelace-Hohlfeld78,Toomre81,Papaloizou-Pringle84,Papaloizou-Pringle85,Hawley87,Lovelace99}. 
The RWI is an ``edge mode'' instability akin to Kelvin-Helmholtz, that converts the extra shear
into vorticity. The large-scale vortices that result are
well-known in the planet formation literature. 

\citet{Barge-Sommeria95}, \citet{Adams-Watkins95}, and \citet{Tanga96}
independently proposed, in the context of primordial disks, that vortices could speed
up planet formation by trapping solids of cm to m
size. The dynamics of this trapping was 
developed in a detailed work by \citet{Chavanis00}, setting much of
the analytical foundations of the
field. \citet{Godon-Livio99,Godon-Livio00} and \citet{Johansen04}
simulated vortices numerically, finding fast trapping of particles but also quick 
dissipation due to (Laplacian) viscosity. These studies, however, did not consider the question of how to
form disk vortices in first place, a question tackled by \cite{Varniere-Tagger06}. These authors show that a
sharp viscosity gradient in the disk leads to a pile-up of
matter, that in turn goes unstable to the RWI. Because the
magnetorotational instability \citep[MRI,][]{Balbus-Hawley91} leads to a significant turbulent
viscosity, \cite{Varniere-Tagger06} suggest that this mechanism could
be at work in the transition between the MRI-active and dead zones. The accumulation of dust
in these self-sustained RWI vortices was subsequently studied by
\citet{Inaba-Barge06}, albeit in the fluid approach, that limited the
dust size they could use. Planetary gap edges were seen to excite
vortices in many simulations in the code-comparison study of
\citet{deValBorro06}, an effect later explained \citep{deValBorro07} in
terms of the RWI as well.

These efforts culminated into a coherent picture of vortex-assisted
planet formation in dead zone vortices by \citet{Lyra08,Lyra09a} and in 
gap edge vortices by \citet{Lyra09b}. These works solved for 
the nonlinear compressible hydrodynamics and the
aerodynamics of interacting particles, demonstrating the
gravitational collapse of the trapped solids, albeit in two
dimensions. The RWI was subsequently studied in barotropic 3D
models by \citet{Meheut10,Meheut12a,Meheut12b}, finding interesting
meridional circulation patterns; in self-gravitating disks with application to 
planet migration in 2D \citep{Lin-Papaloizou11a,Lin-Papaloizou11b,Lin-Papaloizou12}
and 3D \citep{Lin12b}, who find weakening and eventual suppression of the RWI with
increasing disk mass; in MHD by \citet{Lyra-MacLow12}, bringing realism to the dead-zone 
scenario; and by \citet{Lin12, Lin13}, who generalized the linear RWI 
to 3D polytropic and non-barotropic disks, respectively. 

Part of these results have been applied to the field of
transitional disks. The particle size that is
preferentially trapped is set by the friction time, $\tauf$, which is a
function of the gas density and particle radius. A suitable nondimensionalization for the
friction time is the Stokes number, $\St = \varOmega \tauf$,
where $\varOmega$ is the Keplerian frequency. Dust that is too
well-coupled to the gas ($\St\rightarrow 0$) does not suffer friction, and bodies
that are too large ($\St\rightarrow \infty$) have too much inertia to be moved by the
gas: the preferential size for trapping is \St=1 (see e.g. \citealt{Youdin-Goodman05,Youdin08}).
While in the dense, fast rotating, inner regions of primordial disks, 
the preferentially trapped particle size corresponds to meter-size, in
the thin, slowly rotating, outer regions of transitional
disks, the size corresponding to \St=1 drops by about three orders of
magnitude \citep{Brauer07,Pinilla12a}. The resulting trapping of sub-mm and mm-size dust may not
lead to the critical densities necessary to form planets, but they may
well explain the puzzling observed lopsided asymmetries. While the 
motivation and particle sizes are different, the relevant physics 
is scale-free, and thus identical as long as gravity is not involved. 

This property was invoked by \citet{Regaly12} to suggest that the 
sub-mm observations of \citet{Brown09} could be the result of dust
trapping in Rossby vortices. If indeed that is the case, then, 
as the drag force drives dust toward the vortex center, diffusion is
needed to maintain a steady state over the lifetime of the disk
\citep{Klahr-Henning97,Chavanis00}. \citet{Birnstiel13} presented a semi-analytical model that solves for the azimuthal dust 
distribution while using fits from numerical simulations
\citep{Pinilla12b} to constrain the radial morphology. In this work we present a fully analytical 
model for the steady state distribution of dust trapped in vortices, 
accurate to first order in Stokes number, and general in space. In 
\sect{sect:model-equations} we derive the
advective-diffusive equation, and in \sect{sect:coordinate-transformation} the appropriate coordinate
transformation. In \sect{sect:axisymmetric} we solve the equation for
the ``axisymmetric'' case in that coordinate system, and in
\sect{sect:nonaxisymmetric}  we generalize it for
non-axisymmetry. In \sect{sect:observables} 
we derive observational predictions, and apply the model to the Oph IRS 48 system.

\section{Dust steady state} 
\label{sect:model-equations}

Considering the dust is of small sizes, we can treat it as a
fluid. The dust should then follow the continuity equation 

\beq
  \pderiv{\rho_d}{t} = -(\v{w}\cdot\del)\rho_d - \rho_d \Div{\v{w}} - \Div{J}
  \label{eq:continuity-w}
\eeq

\noindent where $\rho_d$ is the dust density, $\v{w}$ is the dust
velocity, and $J$ is the diffusion flux. We take it to be
 
\beq 
J=-D\,\rho_g\grad{\left(\rho_d/\rho_g\right)}\label{eq:j-flux}
\eeq 

\noindent as in the contaminant equation
\citep{Morfill-Volk84,Clarke-Pringle88,Charnoz11}, where $D$ is the  
diffusion coefficient (the diffusion is due
to elliptical turbulence in the vortex core and in general will be
different than the turbulent viscosity in the disk), and $\rho_g$ is
the gas density. We assume that $D$ is constant. A list of the mathematical symbols used in this work, together
with their definitions, is provided in Table~\ref{table:symbols}. 

To derive the velocities, instead of solving the momentum equations
for the dust, we make use of the relative velocity, following
\citet[see also \citealt{Youdin08}]{Youdin-Goodman05}

\beq
\v{w} = \v{u} + \tauf  \grad{h}, 
\label{eq:w}
\eeq
\noindent where $\v{u}$ is the gas velocity. \eq{eq:w} is 
accurate to first order in friction time $\tauf$, assumed 
constant. For isentropic gas, the enthalpy $h$ is defined as $dh = dp /\rho_g$, where $p$ 
is the pressure. As noted by
\cite{Charnoz11}, \eq{eq:continuity-w} can be written as a typical continuity equation with Laplacian diffusion 
\beq
\partial_t \rho_d = -(\v{v} \cdot\del)\rho_d - \rho_d \Div{\v{v}} + D\Laplace\rho_d \label{eq:continuity}
\eeq
\noindent provided that the effective velocity $\v{v}$ is 
\beq
\v{v} \equiv \v{w}  + D\grad\ln \rho_g\label{eq:v-prime}.
\eeq\noindent For isothermal gas the extra term is $D/c_s^2 \grad{h}$,
and, comparing with \eq{eq:w}, its effect amounts to redefining the
friction time as 
\beq
\tau \equiv \tauf + \frac{D}{c_s^2}\label{eq:tau-prime}
\eeq
\noindent combining \eq{eq:w}, \eq{eq:v-prime} and \eq{eq:tau-prime},
we can thus write
\beq
\v{v} = \v{u} + \tau \grad{h}, \label{eq:v}
\eeq \noindent valid for isothermal gas only.

Inside the vortex, the gas flow is divergenceless, and 
we adopt the following model for $\v{u}$
\beq
  u_x = \varOmega_V y / \chi \qquad  u_y= -\varOmega_V x \chi,
  \label{eq:vortex}
\eeq
\noindent where $\chi > 1$ is the vortex aspect ratio (it has
semi-minor axis $a$ and semimajor axis $a\chi$). Notice that the flow eventually
  gets supersonic for large values of $x$ and $y$. This will limit the
  validity of the solution, as the vortices shock beyond the sonic
  perimeter. This effectively leads to a vortex ``boundary'', beyond
  which the motion resumes to the background Keplerian flow.

In this work we consider the Kida solution \citep{Kida81}

\beq
\varOmega_V = \frac{3\varOmega}{2(\chi-1)},  
\eeq
which smoothly matches the above velocity field to the Keplerian
shear; as well as the GNG solution \citep{Goodman87}, that exactly solves the 
compressible Euler equations 

\beq
\varOmega_V=\varOmega\sqrt{3/(\chi^2-1)}.
\eeq

We comment that these solutions make use of the shearing box
  equations, and are thus subject to the same limitations as that
  approximation \citep{Regev-Umurhan08}. In
  particular, the shearing box does not have a radial vorticity
  gradient, and thus cannot excite the RWI
  \citep{Tagger01}. Nevertheless, independently of the excitation
  mechanism, these solutions are good local descriptions of the perturbed
  flow. The GNG solution was used to model vortices
  found in non-linear hydrodynamic global simulations of the
  Papaloizou-Pringle instability \citep{Hawley87}, which is similar to
  the RWI. Recently, \cite{Lin-Papaloizou11a} found that, in
  quasi-steady state, the RWI vortices excited at planetary gap edges 
  resemble vortices formed by perturbing the disk with
  the Kida solution. We are thus confident that the above solutions are suitable
  as a first model for disk vortices.  Moreover, it is straight forward
  to generalize the solutions below to any flow in the form
  $u_x\propto y$ and $u_y\propto -x$.

We note that the dust velocity \eqp{eq:v} is comprised of a divergent-free part, $\v{u}$, and a
curl-free part, $\tau\grad{h}$. The vortex flow attempts to keep the
dust particles on closed elliptic streamlines via $\v{u}$, while friction
attempts to concentrate dust toward pressure maximum via $\tau\grad{h}$. The only effect that attempts to spread out the dust is diffusion
via $D$.  

Taking the divergence of
\eq{eq:v} gives 

\beq
\Div{\v{v}} = \tau \Laplace{h}, 
\label{eq:divv}
\eeq

\noindent and we can find the Laplacian of the enthalpy via the Euler
equation. Adopting the shearing sheet approximation, in steady state
the force balance yields  

\begin{eqnarray}
\pderiv{h}{x} &=& 3\varOmega^2 x + 2\varOmega u_y -
u_y\pderiv{u_x}{y} \nonumber \\
&=& \left(3\varOmega^2 - 2\varOmega\varOmega_V \chi + \varOmega_V^2\right) x
= -\frac{C_1}{\tau} \  x,  \label{eq:gas_mom1}\\
\pderiv{h}{y} &=& - 2\varOmega u_x -
u_x\pderiv{u_y}{x} \nonumber \\
&=& \left(-2\varOmega\varOmega_V/\chi + \varOmega_V^2\right) y = -\frac{C_2}{\tau} \  y.\label{eq:gas_mom2}
\end{eqnarray}

\noindent  Substituting the equations above into \eq{eq:divv}, also
with $\omega_{_V}=\varOmega_V/\varOmega$, the divergence becomes 

\beqn
\Div{\v{v}} &=& -(C_1+C_2) = - C \label{eq:const-div}\\
&=& - \tau\varOmega^2
\left[2\omega_{_V}\left(\frac{\chi^2+1}{\chi}\right) - (2\omega_{_V}^2
  + 3) \right], 
\label{eq:scale-div}
\eeqn

\noindent where we define $C$ as positive, so that the divergence is
negative (physically meaning that the dust gets trapped). Replacing
\eq{eq:const-div} in the modified continuity equation \eqp{eq:continuity}, and setting $\partial_t$ =
0 for steady state, 

\beq
\left(D\Laplace{} -  \v{v}\cdot\del  + C\right)\rho_d = 0. 
\label{eq:steady}
\eeq

Substituting the gas velocity \eqp{eq:vortex}, and dividing by $D$, we
arrive at the modified advection-diffusion equation that should
determine the steady-state distribution of the vortex-trapped dust, 

\beq
\left[\Laplace{} - \left(Ay\chi^{-1} - B_1x\right) \partial_x  +
  \left(A x \chi + B_2y\right) \partial_y + B \right] \rho_d   = 0,  
\label{eq:dust-trapping-cartesian}
\eeq
\\
\noindent where we also substituted  $A=\varOmega_V/D$ and $B_i=C_i/D$.

\begin{table}
\caption[]{Symbols used in this work}
\label{table:symbols}
\begin{center}
\begin{tabular}{l l l}\hline
Symbol & Definition & Description \\\hline
$\tauf$ && friction time\\
$D$ & & dust diffusion coefficient \\
$c_s$ & & sound speed \\
$\tau$ & =$\tauf + D/c_s^2$& effective friction time\\
$\varOmega$ & & Keplerian angular frequency \\
\St & $= \varOmega \tauf$ & Stokes number \\
$t$ &  & time \\
$\rho_g$, $\rho_d$  & & gas and dust density\\
$\v{u}$, $\v{w}$ & & gas and dust velocity\\
$\v{v}$ & $= \v{w} + D\grad\ln\rho$ & effective dust velocity \\
$p$ && gas pressure \\
$h$ &$dh=dp/\rho_g$ & gas enthalpy\\
$\chi$ & & vortex aspect ratio ($>1$) \\
$a$ & & vortex semi-minor axis \\
$\varOmega_V$ & & vortex angular frequency \\
$\omega_{_V}$ &  $=\varOmega_V$/$\varOmega$ & dimensionless vortex frequency  \\
$C$ & $=-\Div{\v{v}}$ &  \\
$A$ & $=\varOmega_V/D$ & \\
$B$  &$=C/D$ & \\
$\nu$ & & azimuth in vortex reference frame\\
$\xi_\pm$ & $ = 1\pm\chi^{-2}$ \\
$H$ & $c_s/\varOmega$ & sonic scale, gas scale height\\
$\delta$ & $D=\delta c_s H$ & dimensionless diffusion parameter\\
$f(\chi)$ & \eq{eq:scale-function} & scale function \\
$S$ & $= \St/\delta$ & dimensionless number \\
$H_g$ & $=H/f(\chi)$ & gas vortex scale length \\
$H_V$ & $=H_g \sqrt{\frac{1}{S+1}}$ & dusty vortex scale length \\
$k$ & $=\sqrt{2}/H_V$ & \\ 
$\zeta$ & $=ka$ & \\
$\tilchi$ & $=\frac{\chi^2-1}{2(\chi^2+1)}$ & \\
$\beta$ & $=(B_1-B_2)/4B$& \\
$k_m$ & $= 1+imA/B$ & \\
$\mathcal{A}_m$ $\mathcal{B}_m$ $\mathcal{C}_m$  &\eqss{eq:ops}{eq:opsw} & differential operators \\
$b_m$ & & constants\\
$\epsilon(\zeta)$ & & non-axisymmetric correction \\
$\varepsilon$ & $=\int \rho_d dV/ \int \rho_g dV$ & global dust-to-gas ratio\\
$\rho_0$ & & max gas density, reference density \\
 & & \\ \hline
\end{tabular}
\end{center}
\end{table}

\section{Change of variable}
\label{sect:coordinate-transformation}

We change variables to the coordinate system used in \citet{Chang-Oishi10}

\beqn
  x &=& a \cos\nu, \label{eq:change-x}\\
  y &=& a\chi\sin\nu.  \label{eq:change-y}
\eeqn

The system is not orthogonal, but it has the advantage of matching the
aspect ratio of the ellipses. (In contrast, the elliptic coordinate
system, though orthogonal, describes a system of confocal ellipses of
different aspect ratio, that does not coincide with the geometry of
the problem.) In these coordinates, the transformations are 

\beq
\left[\begin{array}{c}
    \partial_{a}  \\
    \partial_{\nu}
  \end{array}\right] = \vt{A} 
  \left[\begin{array}{c}
      \partial_{x}  \\
      \partial_{y}
    \end{array}\right] 
\quad {\rm and} \quad 
\left[\begin{array}{c}
    \partial_{x}  \\
    \partial_{y}
  \end{array}\right] = \vt{A}^{-1} 
  \left[\begin{array}{c}
      \partial_{a}  \\
      \partial_{\nu}
    \end{array}\right],  
\eeq

\noindent with 

\beq
\vt{A} = \left[\begin{array}{cc}
\pderiv{x}{a}  & \pderiv{y}{a}  \\
\pderiv{x}{\nu}  & \pderiv{y}{\nu} \\
\end{array}\right] = \left[\begin{array}{cc}
\cos\nu  & \chi\sin\nu  \\
-a\sin\nu  & a\chi\cos\nu \\
\end{array}\right]. 
\eeq

\noindent The inverse matrix is 

\beq
\vt{A}^{-1} = \frac{1}{a\chi} \left[\begin{array}{cc}
a\chi\cos\nu  & -\chi\sin\nu  \\
a\sin\nu  & \cos\nu \\
\end{array}\right].  
\eeq

The transformations are therefore

\beqn
\pderiv{}{x} &=& \cos\nu \pderiv{}{a} - \frac{\sin\nu}{a} \pderiv{}{\nu}, \\
\pderiv{}{y} &=& \frac{1}{\chi}\left(\sin\nu \pderiv{}{a} + \frac{\cos\nu}{a} \pderiv{}{\nu} \right),
\eeqn

\noindent and the Laplacian is thus 

\beqn
\Laplace{} &= &\frac{1}{2}\left[ \epsm \cos 2\nu +  \epsp\right] \partial^2_a  \nonumber \\
                &+& \frac{1}{2a^2}\left[ \epsp - \epsm \cos 2\nu\right] \partial^2_\nu \nonumber \\
                &-& \frac{\sin 2\nu}{a}\epsm \partial^2_{a\nu}   \nonumber \\
                &+& \frac{1}{2a}\left[ \epsp - \epsm \cos 2\nu\right] \partial_a \nonumber \\
                &+& \frac{\sin 2\nu}{a^2} \epsm\partial_\nu, \label{eq:laplace}
\eeqn

\noindent with $\xi_{\pm} = (1 \pm \chi^{-2})$.  As for the advection term, we have 

\beqn
\v{v}\cdot\del &=& (\v{u} + \tau \grad h) \cdot \del \nonumber \\
&=& - \left[\varOmega_V - \frac{\sin2\nu}{2}  (C_1 -  C_2)\right]\partial_\nu \nonumber \\
&&- \left( C_1 \cos^2\nu   + C_2\sin^2\nu \right) a \ \partial_a. \label{eq:advection-term}
\eeqn

The dust-trapping equation is therefore 

\beqn 
&& \left\{ \Laplace{} + \left[A - \frac{\sin2\nu}{2}  (B_1 - B_2)\right]\partial_\nu \ +  \right.  \nonumber\\
&& \left. \white{\frac{1}{1}}\left( B_1 \cos^2\nu   + B_2\sin^2\nu
  \right) a \ \partial_a  + B \right\} \rho_d = 0. \label{eq:dust-trapping-uvzero}
\eeqn

\section{``Axisymmetric'' solution}
\label{sect:axisymmetric}

\subsection{Dust distribution}

We now make the assumption that the dust distribution follows, in shape, that of
the gas (we will relax this approximation in the next section). In this case, the
dust distribution follows ellipses of equal aspect ratio. So,
$\partial_\nu$ = 0, ``axisymmetric'' in the ($a,\nu$)
coordinates. \eq{eq:dust-trapping-uvzero} becomes

\beqn
&&\left\{\frac{1}{2}\left(\epsm \cos 2\nu +\epsp\right) \partial^2_a  +  \left[\frac{1}{2a}\left( \epsp - \epsm\cos 2\nu\right) \right.\right. \nonumber\\
&&\left.\left.\white{\frac{1}{1}}+ (B_1\cos^2\nu +  B_2\sin^2\nu)a\right] \partial_a  + B\right\} \rho_d = 0. \label{eq:axi}
\eeqn

We now integrate the above equation in $\nu$, from 0 to 2$\pi$. This yields

\beq\label{eq:dust-trapping-axis}
\left[\partial^2_a  +  \left(\frac{1}{a} +  \frac{k^2}{2}a\right) \partial_a  + k^2\right]\rho_d = 0, 
\eeq

\noindent where we define $k^2=2B/\epsp$. Note that the parameter $A$
is absent because it represents advection by the vortex, which only
move dust particles along the same ellipse, not across it. It is
not relevant in the $\nu$-averaged problem. The solution of
\eq{eq:dust-trapping-axis} is 

\beq
\rho_d(a) = \exp\left(-\frac{k^2a^2}{4}\right)  \left[c_1 + c_2 {\rm
    Ei}\left(\frac{k^2a^2}{4}\right)\right],
\eeq

\noindent where $c_1$ and $c_2$ are constants, and ${\rm Ei}(x)$ is
the exponential integral function. Since it diverges at the origin, $c_2$ has to be zero, and 

\beq
\rho_d(a) = \rho_{d\,{\rm max}} \ \exp\left(-\frac{a^2}{2H_V^2}\right),
\label{eq:gen_axi}
\eeq

\noindent with $H_V = \sqrt{2}/k$ for symmetry with the gas sonic scale. We can 
rewrite this length scale recalling that $k^2=2B/\epsp$ and
$B=C/D$. We can substitute the diffusion coefficient $D=\delta \varOmega H^2$ where 
$\delta$ is a dimensionless coefficient, and $\St = \tauf\varOmega$ for 
the Stokes number, writing thus 

\beqn
\label{eq:k}
k^2 = \frac{2(\St+\delta)}{\delta H^2} f^2(\chi),
\eeqn 

\noindent so

\beq
 H_V = \frac{H}{f(\chi)} \sqrt{\frac{\delta}{\St+\delta}}. 
\label{eq:hv-pre}
\eeq
\noindent following \cite{Jacquet12} we define $S=\St/\delta$. The vortex scale
length is therefore
\beq
H_V = \frac{H}{f(\chi)} \sqrt{\frac{1}{S+1}}
\label{eq:hv}
\eeq

\begin{figure}
  \begin{center}
    \resizebox{\columnwidth}{!}{\includegraphics{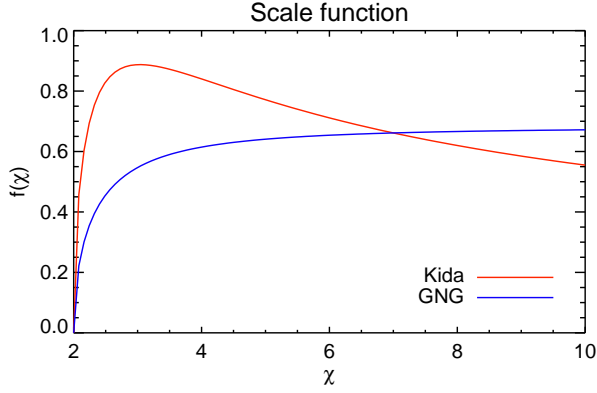}}
  \end{center}
\caption[]{The scale function $f(\chi)$, defined by
  \eq{eq:scale-function}, for the Kida ($\varOmega_V=3/2
  \ \varOmega_K/(\chi-1)$) and GNG ($\varOmega_V=\varOmega_K
  \sqrt{3/(\chi^2-1)}$) solutions, respectively. The scale function is
  related to the square root of the negative of the divergence
  \eqp{eq:scale-div}, and defined only for $\chi>2$. For smaller $\chi$ the
  divergence flips positive, meaning that dust is expelled from the
  vortex instead of getting trapped. This happens because of the
  correlation between $\varOmega_V$ and $\chi$. The aspect ratio shrinks
  as the vortex intensifies. At some point, the vortex rotates too
  fast, and particles are expelled by the centrifugal force.}
 \label{fig:scale-function}
\end{figure}

\noindent In these equations, the scale function $f(\chi)$ is given by 

\beqn
f^2(\chi) &=& \epsp^{-1} \left[2\omega_{_V}\left(\frac{\chi^2+1}{\chi}\right) - (2\omega_{_V}^2 + 3) \right]\nonumber \\
          &=& 2\omega_{_V}\chi - \epsp^{-1}(2\omega_{_V}^2 + 3),
\label{eq:scale-function}
\eeqn

\noindent and depends on the vortex solution
used. We plot $f(\chi)$ for the Kida and GNG solutions in
\fig{fig:scale-function}. They are defined in the real axis only for $\chi > 2$ ($f^2
< 0$ for $0 < \chi < 2$ ). The Goodman solution tends to an asymptote
around 0.7. The Kida solution has a  tail around $0.5\pm0.25$ in the
interval of physical relevance ($2 < \chi \lesssim 10$). 

\begin{figure}
\begin{center}
  \resizebox{\columnwidth}{!}{\includegraphics{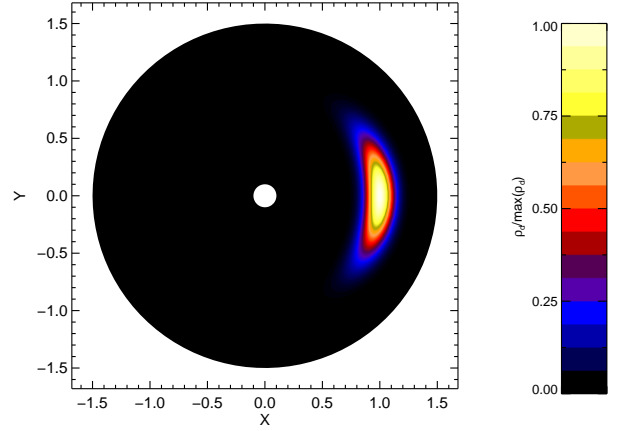}}
 \end{center}
\caption[]{Three parameters, plus a vortex solution, control the dust distribution. 
The figure shows the appearance of the dust trapped in a Kida vortex of $\chi=4$, 
for $S$=1, in a disk of aspect ratio $H/r$=0.1.} 
 \label{fig:disk}
\end{figure}

We show in \fig{fig:disk}, in the inertial frame, 
the dust distribution for $S$=1 in a Kida vortex of $\chi=4$ embedded in a 
disk of aspect ratio $H/r$=0.1, where $r$ is the stellocentric
distance. We caution that this image extrapolates the spatial
range of applicability of the shearing box approximation used to construct
the solution.

It is worth noting that for certain vortex models and/or
aspect-ratios, the Gaussian solution, \eq{eq:gen_axi}, is in fact an 
exact solution to the dust-steady state equation, \eq{eq:dust-trapping-uvzero}.  
We will explore this in more detail  in \sect{sect:nonaxisymmetric}, but one can check this by inserting
\eq{eq:gen_axi} into \eq{eq:dust-trapping-uvzero}, and finding the condition 
for the coefficient of the trigonometric terms to vanish. 
In this special case, explicitly averaging over $\nu$ is not
required to remove the $\nu$-dependence from the 
problem.

\subsection{Gas distribution}

\eq{eq:gen_axi} allows us to calculate the gas distribution. For
  that we recall that  for tracer particles ($\St=0$), the dust distribution should
  mimic that of the gas. The distribution should thus be 
\beq
\rho_g(a) = \rho_{g\,{\rm max}} \ \exp\left(-\frac{a^2}{2H_g^2}\right), \label{eq:gas_axi}
\eeq
\noindent with 
\beq
H_g = {H_V}{\vert_{\St=0}} = H/f(\chi) \label{eq:hg}
\eeq \noindent and
$\rho_{g\,{\rm max}}$, the maximum gas density{\footnote{Note that 
\eq{eq:gas_axi} is the gas density averaged over $\nu$ at fixed $a$.  One may  
directly integrate the gas momenta equations to see that the gas
density/pressure depends, in general, on both $a$ and $\nu$.}}.

Notice that for $\St=0$ the effect of diffusion cancels out. This is because
the diffusion is proportional to the gradient of the dust-to-gas
ratio \eqp{eq:j-flux}, which is zero for tracer
particles.

\begin{figure}
  \begin{center}
    \resizebox{\columnwidth}{!}{\includegraphics{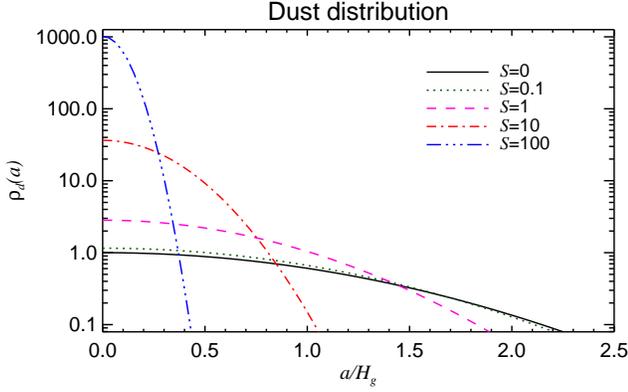}}
  \end{center}
\caption[]{Dust distribution for the ``axisymmetric'' case (in
    the coordinate system defined by \eqs{eq:change-x}{eq:change-y}). The
    maximum density is proportional to $(S+1)^{3/2}$. Curves for
  $S$=0, 0.1, 1, 10, and 100 are shown. The
  $S$=0 case represents tracer particles and, consequently, the gas
  density. The $x$-axis is $a/H_g$, where $H_g=H/f(\chi)$ is the
  vortex scale length in the gas phase, with $H$ the sonic scale and
  $f(\chi)$ the model-dependent scale function
  \eqp{eq:scale-function}.}
 \label{fig:gaussian}
\end{figure}

\section{Non-axisymmetric corrections}
\label{sect:nonaxisymmetric} 

We now consider the non-axisymmetric problem ($\partial_\nu\neq0$). 
We explicitly show that such effects are small 
in the vortex core provided the effective Stokes number $\overline{\mathrm{St}}\equiv\mathrm{St}+\delta$ 
is not large. These requirements will become
apparent as we proceed through the solution method. In this section we
consistently refer to ``axisymmetric'' as $\nu$-symmetry in the coordinate
system defined by \eqs{eq:change-x}{eq:change-y}.

\subsection{Conversion to ordinary differential equations}

The dust density $\rho_d$ is periodic in the $\nu$ co-ordinate. We 
therefore seek solutions of the form

\beq\label{eq:series-n}
\rho_d(a,\nu) = {\rm Re}
\left[\sum_{n=0}^\infty\rho_n(a)\exp{\left(\mathrm{i}n\nu\right)} \right].
\eeq

For convenience, we will drop the real part notation from now on. Inserting
\eq{eq:series-n} into the partial differential equation \eqp{eq:dust-trapping-uvzero},
multiplying by $\exp{(-\mathrm{i}m\nu)}$, and integrating the resulting
expressions over the $\nu$ co-ordinate, we arrive at a set of coupled
ordinary differential equations, 

\beq\label{eq:ode2}
\mathcal{B}_m\rho_{m-2} (\zeta)+ \mathcal{A}_m\rho_m(\zeta) + \mathcal{C}_m\rho_{m+2}(\zeta) = 0,
\eeq

\noindent where $\zeta\equiv ka$, and

\beqn
\mathcal{B}_m &\equiv& \tilchi\frac{d^2}{d\zeta^2} + \left[\beta \zeta-
  \frac{2\tilchi}{\zeta}
  \left(m-\frac{3}{2}\right)\right]\frac{d}{d\zeta}+\left(m-2\right)\left(\frac{m\tilchi}{\zeta^2}-\beta\right), \notag\\
&&\label{eq:ops}\\ 
\mathcal{A}_m&\equiv&\frac{d^2}{d\zeta^2} +\left(\frac{1}{\zeta}+\frac{\zeta}{2}\right)\frac{d}{d\zeta} +\left(k_m^2-\frac{m^2}{\zeta^2}\right),\\
&&\notag\\
\mathcal{C}_m &\equiv& \tilchi\frac{d^2}{d\zeta^2}
+\left[\beta \zeta +
  \frac{2\tilchi}{\zeta}\left(m+\frac{3}{2}\right)\right]\frac{d}{d\zeta}
+ \left(m+2\right)\left(\frac{m\tilchi}{\zeta^2}
  +\beta\right),\notag\\
&&\label{eq:opsw}
\eeqn
where $\tilchi\equiv(\chi^2-1)/[2(\chi^2+1)]$, $k_m^2 \equiv 1+\mathrm{i}mA/B$, and
\beq
\beta \equiv \frac{B_1-B_2}{2k^2(1+\chi^{-2})} = \frac{B_1 - B_2}{4B}.
\eeq

Note that $\beta$ is a function of the aspect-ratio depending on the vortex model.
\eq{eq:ode2} holds for each $m$ except for $m=0$ for which the $\rho_{m-2}$ terms are absent. Each 
$\rho_m$ couples to $\rho_{m\pm2}$ through operators $\mathcal{B}_m$
and $\mathcal{C}_m$. The axisymmetric problem is recovered by
setting $\rho_{m>0}=0$. 

We expect $\rho_d(a,\nu)$ to have even symmetry in
$\nu$ because of the elliptical nature of the vortex streamlines. 
Henceforth we only consider even $m$. We seek solutions with  
$\rho_m^\prime(0)=0$ (where the prime denotes derivative with respect to the argument) and $\rho_{m\geq2}(0)=0$, so that
$\partial_x\rho_d=\partial_y\rho_d=0$ at the origin, consistent with 
dust reaching maximal density there.  

\subsection{Operator properties}

Consider 
\begin{align}
	g_m(\zeta) \equiv \zeta^m\exp{(-\zeta^2/4)}.
\end{align}
Then we find that 
\beqn
\mathcal{B}_mg_{m-2} &=&\frac{1}{4}\left(\tilchi - 2\beta\right)g_m,\\
\mathcal{A}_mg_m &=& \left(k_m^2 - \frac{m}{2} - 1\right)g_m,\\
\mathcal{C}_mg_{m+2}&=& \left[4\tilchi(m+1)(m+2) +2(\beta-\tilchi)(m+2)\zeta^2 \right.
\notag\\ &+&\left. \frac{\left(\tilchi - 2\beta\right)}{4}\zeta^4\right]g_m. 
\eeqn
The first two expressions will be useful in constructing nearly-axisymmetric solutions. 

\subsection{Exact axisymmetric solutions}
It is useful to see how the formulation above connects with the  axisymmetric solutions discussed in the previous section.  Consider the special case where $\tilchi = 2\beta$, so that 
$\mathcal{B}_mg_{m-2} = 0.$
Then the complete solution to \eq{eq:ode2} is $\rho_0 = b_0e^{-\zeta^2/4}$ with $\rho_{m>0} \equiv 0$, and $b_0$ is an arbitrary constant. That is, if $\tilchi=2\beta$ then the dust distribution is exactly axisymmetric. 

\subsubsection{Dust in a GNG vortex is axisymmetric}
For the GNG vortex, one can verify that $\tilchi\equiv 2\beta$, implying dust density only depends on the ellipse under consideration, not the position along it. This is because the GNG vortex has no pressure gradient along the elliptical streamlines \citep{Chang-Oishi10}. 

\subsubsection{Condition for dust in a Kida vortex to be axisymmetric}
For the Keplerian Kida vortex, we find
\begin{align}
\tilchi - 2\beta = \frac{\chi(\chi-1)(\chi-7)}{2(\chi-2)(2\chi+1)(\chi^2+1)}.
\end{align}
The dust distribution is exactly axisymmetric for aspect-ratio $\chi=7$, which is also when the Keplerian Kida vortex has no pressure gradient along its elliptical streamlines \citep{Chang-Oishi10}.

\subsection{Source term approximation}
In preparation for constructing non-axisymmetric solutions, we here describe the \emph{source term approximation} \citep{Zhang06}.
We assume that $|\rho_m|$ decreases with $m$, so that in \eq{eq:ode2} the $\mathcal{C}_m\rho_{m+2}$ term has smallest magnitude.  Neglecting it as a first approximation, we solve 

\begin{align}
\mathcal{A}_m\rho_m = \begin{cases}
        0 & m =0 \\
	-\mathcal{B}_m\rho_{m-2} & m \geq 2.
\end{cases}
\end{align}
The solutions are 
\begin{align}\label{eq:series1}
\rho_m (\zeta)= b_m g_m(\zeta),
\end{align}
with
\begin{align}
b_m  = -\frac{\left(\tilchi - 2\beta\right)}{2\left[2k_m^2 - (m+2)\right]}b_{m-2}
\end{align}
for $m\geq2$, and $b_0$ is arbitrary as before. Note that $b_m=0$ for odd $m$ because $b_1=0$ since 
we require $\rho_1^\prime(0)=0$. Then, by induction 
\begin{align}\label{eq:induction}
b_m  = \left(-1\right)^{m/2}\frac{\left(\tilchi/2-\beta\right)^{m/2}}{\prod_{l=1}^{m/2}
\left(2k_{2l}^2 -2l - 2\right)}b_0,
\end{align}
for even $m\geq 2$.  

The source term approximation assumes 
$R\equiv|\mathcal{C}_m\rho_{m+2}|/|\mathcal{A}_m\rho_m| \ll 1.$ 
For given $\zeta$, the solution $\rho_m=b_mg_m$ is consistent with this requirement if $|k_m^2|\gg1$, corresponding to small effective Stokes number.  
 However, this approximation will eventually fail for large $\zeta$ because the solution above implies $R\propto \zeta^4$ for $\zeta\gg1$. Thus the solution is only self-consistent for sufficiently small $\zeta$ and/or $\overline{\St}$. Nevertheless,
we comment that the closed-formed solutions obtained here may be useful in
an iterative scheme to obtain numerical solutions to the full set of ODE's. 

\subsection{Weakly non-axisymmetric dust distributions}
We are now ready to construct non-axisymmetric solutions. 
Consider a Keplerian Kida vortex with $\chi\neq 7$, meaning that the effective 
frictional force on the dust has a non-vanishing component along the
fluid velocity vector. (I.e. dust particles are accelerated along the
ellipse.) We assume non-axisymmetry in the dust distribution is
sufficiently weak, so one may truncate the series solution at
$m=2$. Thus we set $\rho_{m>2}\equiv0$. Let
\begin{align}
\rho_0(\zeta) = b_0 g_0(\zeta) + \epsilon(\zeta),
\end{align}
where $\epsilon(x)$ represents the correction to the axisymmetric
solution due to $\rho_2(\zeta)$. The ODEs to be solved are

\begin{align}
&\mathcal{A}_0\epsilon(\zeta) = -\mathcal{C}_0\rho_2(\zeta),\label{eq:m2_eq1}\\
&\mathcal{A}_2\rho_2(\zeta)   = -\mathcal{B}_2\left(b_0g_0 + \epsilon\right).\label{eq:m2_eq2}
\end{align}
To make further progress, at this stage we \emph{assume} that the
$\epsilon$ term in Eq. \ref{eq:m2_eq2} can be neglected, so $\rho_2 = b_2 g_2$ with $b_2$ given by the source term approximation. 
This means that  

\beq\label{eq:rho2}
\frac{\rho_2}{\rho_0} = \frac{b_2}{b_0}\zeta^2,
\eeq

\noindent implying non-axisymmetry becomes significant for sufficiently large $\zeta$, 
and truncating the series at $m=2$ is no longer self-consistent. 
However, in practice the ratio $|b_2/b_0|$ is small. 
For example, inserting $\chi=4$ gives $|b_2/b_0|\simeq0.1\%$ for
$\overline{\St}=0.1$ and $|b_2/b_0|\sim 1\%$ for $\overline{\St}=1$. Since
most of the dust is contained within $\zeta\lesssim 1$, we conclude that
non-axisymmetry is {\it in general} a small effect.

We can use \eq{eq:rho2} in \eq{eq:m2_eq1} to
calculate the correction term $\epsilon$. We find

\beq
\epsilon(\zeta) = \frac{1}{8}b_2g_2\left[-16\tilchi+\left(\tilchi - 2\beta\right)\zeta^2\right].
\eeq

Collecting the above results and Taylor-expanding the $g_m$'s, our
weakly non-axisymmetric solution for $\zeta\ll 1$ reads:

\begin{align}
&\rho_0(\zeta) =1 -  \frac{\zeta^2}{4}\left[1-\frac{\tilchi\left(\tilchi - 2\beta\right)}{\mathrm{i}A/B - 1/2}\right]+ O(\zeta^4),\\
&\rho_2(\zeta) =-\frac{\left(\tilchi - 2\beta\right)}{2\left(4\mathrm{i}A/B - 2\right)}\zeta^2+O(\zeta^4)
\end{align}
where we have used the definition of $k_m$ and set $b_0=1$
  without loss of generality. In the previous section, we obtained the axisymmetric solution
assuming the non-axisymmetric components are negligible. 
Here, we see explicitly that the axisymmetric solution in fact leads
to non-axisymmetry through the coupling terms, but these corrections
are small for $\overline{\St} \ll 1$, because $B \propto \overline{\St}$. We conclude that dust in the vortex core is effectively axisymmetric.

\subsubsection{Consistency check}
Using the above expression for $\epsilon(\zeta)$, we can evaluate
$\mathcal{B}_2\epsilon(\zeta)$ in order to assess our assumption that
$\epsilon(\zeta)$ has a negligible contribution to $\rho_2$. We find

\beqn
\mathcal{B}_2\epsilon(\zeta) =&& \left[32\tilchi(5\tilchi - 6\beta)-16(\tilchi-2\beta)(2\tilchi-\beta)\zeta^2
\right.\notag\\&&\left.+(\tilchi - 2\beta)^2\zeta^4\right]\frac{b_2g_2}{32}.
\eeqn

Provided that $|k_2^2|\gg1$ and $\zeta$ is not large,  
this term is indeed small compared to the first term on the RHS of \eq{eq:m2_eq2}. For example, considering $\zeta=1$, 
for $\chi=4$ and $\overline{\St}=0.1$ we obtain $|\mathcal{B}_2\epsilon|/|\mathcal{B}_2\rho_0|\simeq0.02$. 
Even with $\overline{\St}=1$, this ratio $\sim0.2$ is not large. We
conclude that our solution procedure above is self-consistent.

\section{Observational predictions}
\label{sect:observables}

Having arrived at the ``axisymmetric'' solutions (in the
  $a$-$\nu$ plane, \sect{sect:axisymmetric}), and shown that deviations from
  $\nu$-symmetry are small (\sect{sect:nonaxisymmetric}), we go back to
  the solutions of \sect{sect:axisymmetric} to derive observational predictions.

\subsection{Dust - gas contrast}

\eq{eq:gen_axi} and \eq{eq:gas_axi} also allows us to calculate the gas-dust
density contrast, and, therefore, $\rho_{d\,{\rm max}}$ as a function of
$\rho_{g\,{\rm max}}$. For that, we calculate the volume integral of $\rho_d$
and $\rho_g$. These, in turn, need the dependencies on the vertical
coordinates $z$. These are straightforward, being $\exp(-z^2/2H^2)$ for the gas and
$\exp(-z^2/2H_d^2)$ for the dust, with $H_d=H/\sqrt{(1+S)}$ \citep{Dubrulle95}. Integrated over plus and minus infinity, these yield $\sqrt{2\pi}H$ and
$\sqrt{2\pi}H_d$, respectively. We have thus

\beqn
\int\rho_d(a,z) dV  &=& \rho_{d\,{\rm max}} \ \frac{(2\pi)^{3/2}}{\sqrt{S+1}} H \int_0^\infty {\rm  e}^{-a^2/2H_g^2 \ (S+1)} \ a\chi \ da \notag\\
&&= \rho_{d\,{\rm max}} \ \left(\frac{2\pi}{S+1}\right)^{3/2} \chi H H_g^2, \label{eq:integral-dust}\\ 
\int\rho_g(a,z) dV  &=& \rho_{g\,{\rm max}} \ (2\pi)^{3/2} H \int_0^\infty {\rm e}^{-a^2/2H_g^2} \ a\chi \ da\ \notag\\
 &&= \rho_{g\,{\rm max}} \ (2\pi)^{3/2} \chi H H_g^2. \label{eq:integral-gas}\\
\eeqn 
\noindent Dividing \eq{eq:integral-dust} by \eq{eq:integral-gas}, the
ratio of the integrals in the left hand sides is the global dust-to-gas
ratio, $\varepsilon$. The density enhancement factor is thus
\beq
  \rho_{d\,{\rm max}} =  \varepsilon \ \rho_0 \ (S+1)^{3/2} \label{eq:maxrhod}
\eeq where $\rho_0=\rho_{g\,{\rm max}}$ is an appropriate reference
density. The full expression for the dust density is therefore

\beq
   \rho_d(a,z) = \varepsilon \, \rho_0 \, (S+1)^{3/2} \  \exp{\left\{-\frac{\left[a^2f^2(\chi) + z^2\right]}{2H^2}(S+1)\right\}} 
\label{eq:sect6-exp}
\eeq

\eq{eq:maxrhod} shows that the dust-to-gas ratio at the origin (vortex center)
is related to the total dust-to-gas mass ratio by a simple function of $S$.  
In this enhancement, only a third (in log) is caused by sedimentation. The
rest is due to in-plane vortex capturing. Midplane dust
distributions for different values of $S$ are plotted in \fig{fig:gaussian}, as a 
function of $a/H_g$. 

\subsection{Trapped mass}

For the total trapped mass, we simply need
to integrate \eq{eq:sect6-exp}, which amounts to replacing
\eq{eq:maxrhod} in \eq{eq:integral-dust}
 \beq
\int\rho_d(a,z) dV  = \left(2\pi\right)^{3/2} \, \varepsilon \, \rho_0 \ \chi H H_g^2 \label{eq:dust-mass}\\ 
\eeq

\subsection{Dust density contrast}

The contrast in the same orbit is found by calculating the
  minimum dust density and comparing it to \eq{eq:maxrhod}. By substituting the gas solution \eqp{eq:gas_axi}
  into \eq{eq:gen_axi} we can write
\beq
\frac{\rho_{d\,{\rm max}}}{\rho_{d\,{\rm min}}} = \frac{\rho_{g\,{\rm max}}}{\rho_{g\,{\rm min}}} \exp{(S)},\label{eq:max-contrast}
\eeq
which is the same result
as found by \cite{Birnstiel13}, provided a suitable choice is made for
$\delta$ (we do not assume a relationship between $\delta$ and $\alpha$
because the turbulence in the vortex core is locally generated and
unrelated to the disk turbulence, c.f., elliptic instability,
\citealt{Lesur-Papaloizou10,Lyra-Klahr11}).
 The minimum densities occur at the boundary of the vortex, which is
 the sonic perimeter where shocks occur. Its limit is found by
 writing the vortex velocity \eqp{eq:vortex} as a Mach number 
\beq
  {\rm Ma} = \frac{|u_y|}{c_s} = \omega_V\chi \  \frac{x}{H} 
  \label{eq:vortex-mach}
\eeq and setting ${\rm Ma} = 1$. This yields the
boundary at 
\beq
a_s = H (\chi \omega_V)^{-1} \label{eq:sonic}
\eeq where the subscript $s$ stands for
sonic. The Kida solution asymptotically reaches $a_s=2H/3$, while the
GNG solution asymptotically reaches $a_s=H/\sqrt{3}$. In the physical
range of relevance ($2\lesssim \chi \lesssim 10$), they both yield values around $H/2$, which matches  
the results of numerical simulations. Substituting \eq{eq:sonic} in \eq{eq:gas_axi}, the gas density contrast is 
\beq
  \frac{\rho_{g\,{\rm max}}}{\rho_{g\,{\rm min}}} =
  \exp{\left[\frac{f^2(\chi)}{2\chi^2\omega_V^2}\right]}, \label{eq:gas-contrast}
\eeq
\noindent For neither the Kida nor the GNG solutions does this quantity
deviate much from unity. This is because the argument in the exponent tends
asymptotically in both cases to small fractions of $f^2$; $2/9$ in the
Kida case, $1/6$ in the GNG case. 

\subsection{Measuring $\delta$}

Closed elliptic streamlines are subject to the elliptic
instability, which leads to subsonic turbulence in the vortex core 
\citep{Lesur-Papaloizou10,Lyra-Klahr11}. To directly measure $\delta$, 
the turbulent diffusion parameter, one would need to measure the 
turbulent velocity field. As $\alpha$, the Shakura-Sunyaev viscosity
parameter \citep{Shakura-Sunyaev73}, $\delta$ can be
defined as the ratio of stress over pressure. If the turbulence is
isotropic in the midplane, one can write 
\beq
\delta = v_{\rm  rms}^2/c_s^2,
\eeq where $v_{\rm rms}$ is the rms of the turbulent velocities. 
The beam smearing would render the velocity field 
unresolved even for moderately close systems, so one should 
look for unresolved signatures. Spectroscopically, this extra rms velocity should have 
an effect similar to microturbulence, providing a slight extra
broadening to the Doppler core of suitable spectral lines.

For gas temperatures ranging 20-200\,K, assuming that the gas is a 5:2 hydrogen to
helium mixture (mean molecular weight of 2.4), the isothermal sound
speeds range 0.26-0.83 \, km/s. Considering that typical velocities of subsonic
turbulence are $\approx$10\% of the sound speed ($\delta \approx
10^{-2}$), the typical velocity signal for 200\,K would be of the order
of $\leq$0.1\,km/s. As \cite{vanderMarel13} quote a
sensitivity limit of 0.2\,km/s for their ALMA observations of
Oph IRS 48, only the $\geq$2$\sigma$ tail of the turbulent velocity field
should be detectable.

If a direct determination of $\delta$ does not sound promising,
  an indirect way is possible by measuring $S$ and $\St$. The
  parameter $S$ can be determined via the dust-density
  contrast with \eqs{eq:max-contrast}{eq:gas-contrast}, or via the
  dust-gas contrast at maximum \eqp{eq:maxrhod}. The Stokes number is 
\beq
    \St = \tauf  \varOmega = \sqrt{\frac{\pi}{8}} \
    \frac{a_\bullet}{H} \ \frac{\rho_\bullet}{\rho_g}
\eeq
\noindent where $a_\bullet$ is the particle radius and $\rho_\bullet$ the
particle internal density.

\subsection{Application to Oph IRS 48}
\label{sect:oph-irs-48}

We now apply our model to the observed Oph IRS 48 system, with the
parameters derived by \cite{vanderMarel13}.The dust contrast in the same orbit is 130, which, according to
\eq{eq:max-contrast} and \eq{eq:gas-contrast} for $\chi=3.1$, sets $S=4.79$ and
$S=4.82$ for the Kida and GNG solutions, respectively.  The values are
close because the gas contrast is small \eqp{eq:gas-contrast}.

The dust temperature derived by the authors is 60\,K. Assuming this is the
  same as the gas temperature, and a mean molecular weight of 2.4, the isothermal sound speed is 
  $c_s \approx {\rm 456 \, cm/s}$. At $r_0$=63\,AU, around a $2 M_\sun$ star, this translates into an aspect
  ratio of $H/r \approx0.09$, or $H\approx5.4$ \,AU. As for the
    particle radius, the ALMA data is sensitive up to $a_\bullet\approx1.5$\,mm, and we
    take this size to be representative. 

The gas mass is quoted to range between 19-27 Jupiter masses,
  measured from a ring centered at 60\,AU. The signal-to-noise is too
  low to derive a radial extent, but assuming it ranges 50-70\,AU, the
  gas surface density should range 20-30 \ g\,cm$^2$. We take
  $\varSigma_g$=25\ g\,cm$^2$ as best estimate, which, for the scale
  height derived above, translates into $\rho_g=\varSigma_g/(\sqrt{2\pi}H)
  \approx 1.25\times 10^{-13}\ {\rm g\,cm}^{-3}$.

For particles of material density $\rho_\bullet=0.8 \ {\rm g\,cm}^{-3}$, the
Stokes number should then be $\St \approx 0.008$. For $S=4.8$, this translates
into $\delta \approx 1.5\times10^{-3}$, meaning typical turbulent velocities in the
vortex core at $\sqrt{\delta} \approx 4\%$ of the sound
speed. These velocities fall {\it squarely} within the range expected for the
  elliptic instability \citep{Lesur-Papaloizou10,Lyra-Klahr11},
  that shows a maximum speed of 10\% of the speed of sound.

As for the trapped mass, \citet{vanderMarel13} measures
  9\,$M_\oplus$. For the typical interstellar
dust-to-gas ratio of $\varepsilon=0.01$, \eq{eq:dust-mass} yields 6 
and 17\,$M_\oplus$ for Kida and GNG, respectively. Given the
approximations, assumptions, and uncertainties, the agreement within a
factor 2 is remarkable.

Although these values seem reasonable, it should be noted that
for Oph IRS 48 the candidate planet is at $\approx$20\,AU, whereas the dust trap is at 63\,AU. Even though the
planet is supposed to be massive (planet-to-star mass ratio
$5\times10^{-3}$), gaps are not expected to be that wide. The supposed
vortex also seems to be very big, with a semiminor axis of 17\,AU. For
a temperature of 60\,K, this corresponds to over 3$H$, which is far
from the $\approx H/2$ expected from numerical simulations and
\eq{eq:sonic}. Relaxing the approximation that the gas and dust 
have the same temperature does little to solve the
discrepancy. Because $H \propto c_s \propto \sqrt{T}$, a vortex
six times bigger means a temperature thirty-six times hotter. This would 
bring the gas temperature above 2\,000\,K, which is unrealistic.

\section{Conclusions}

We solve for the distribution of dust trapped in
disk vortices, in steady state between gas drag, that 
tends to drive dust into the vortex, and diffusion, that expels
it. \eqs{eq:gen_axi}{eq:hv}, with coefficient given by \eq{eq:maxrhod},
 are our result for a distribution with
``axis-symmetry''  in the coordinate system defined by
\eqs{eq:change-x}{eq:change-y}. That is, consisting of ellipses of
equal aspect ratio as those of the gas vortex. The solution has some remarkable
properties. It is a Gaussian of standard deviation $H_V$, where, given
the angular velocity $\varOmega_V$ of the vortex, $H_V$
is determined by three quantities. These are: 
the sonic length and gas scale height, $H$; the vortex aspect ratio
$\chi$; and $S=\St/\delta$, the relative strength of drag to
diffusion. The importance of this latter parameter had already been 
hinted upon by \citet{Cuzzi93} and \citet{Dubrulle95} 
in the context of steady states of dust sedimentation, and by
\citet{Klahr-Henning97} for vortices in the meridional plane. An insightful study by 
\citet{Jacquet12} emphasized the relevance of this parameter for 
global redistribution of solids. \citet{Birnstiel13} also find this to be 
the parameter of relevance in their semi-analytical model. 

Transitional disks provide an interesting venue where to test the
model in an astrophysical context, since all three parameters can be
derivable from data. The vortex aspect ratio is readily observable,
and $H$ follows from the temperature ($H=c_s/\varOmega$). The 
parameter $S$ follows from the density contrast (either dust-gas
contrast at maximum or dust constrast in the same
orbit). Disentangling $\St$ from $\delta$ in this parameter requires
directly measuring at least one of these quantities. The diffusion
parameter $\delta$ is in principle not equal to
$\alpha$ (the dimensionless gas viscosity of
\citealt{Shakura-Sunyaev73}), because the processes generating turbulence in the
vortex and in the disk are different. The latter is supposedly the MRI, whereas the former is the elliptic or
magneto-elliptic instability (see \citealt{Lyra13} and references
therein). A direct measure of $\delta$ would require measuring the
velocity field inside the vortex, that would appear spectroscopically 
as a slight extra line broadening. However, this would be
difficult because the signal is too small. Measuring $\St$ requires
knowing the gas density and temperature, the particle radius and
internal density. Of these, the internal density is difficult to measure
directly and should be inferred by laboratory experiments. We apply
the model to the Oph IRS 48 system, finding consistent values. The
Stokes number for the 1.5\,mm particles is estimated at
$\St \approx 0.008$, implying 
$\delta \approx 1.5\times 10^{-3}$, and turbulent velocities in 
agreement with numerical simulations. The total dust masses we estimate 
are within a factor 2 of the measured value. 

We also solve for the non-axisymmetric problem, showing that, for the
vortex core, it is in general but a small correction. The solution is
\eq{eq:series-n}, with ``radial'' basis functions given by
\eq{eq:series1} and coefficients given by \eq{eq:induction}. In practice, the magnitude
of the higher non-axisymmetric modes fall fast as $m$ increases, and 
only the $m=2$ term would provide an appreciable deviation from
$\nu$-symmetry. We find that non-axisymmetry in dust 
is associated with non-zero pressure gradients along elliptical streamlines of the vortex.

We recall that aside from planetary gap edges, self-sustained disk
vortices may also result from either RWI at the boundary between the MRI-active
and dead zones, or convective-like nonlinear baroclinic instabilities \citep{Klahr-Bodenheimer03,Klahr04,Petersen07a,Petersen07b,Lesur-Papaloizou10,Lyra-Klahr11,Raettig13}. 
These processes, however, are not reasonable in the context of outer
regions of transition disks: the outer edge of the dead zone is quite smooth
\citep{Dzyurkevich13,Landry13}, whereas the RWI requires a 
sharp enough transition; as for the baroclinic instability, it
requires finite thermal diffusion, whereas the thin outer disk
  is supposed to radiate efficiently. This leaves gap-edge RWI as the only currently known
plausible mechanism to excite such vortices. However, this interpretation 
is not without difficulties, because, as noted in
\sect{sect:oph-irs-48}, the dust trap is too far out (63\,AU) to be
the result of a gas gap carved by a planet at 20\,AU, and because its
radial size ($\approx$35\,AU) would imply an unrealistic high
gas temperature. Future modeling should aim at solving these discrepancies.

\acknowledgments WL acknowledges financial support by the National Science
Foundation under grant no. AST10-09802. This work was performed in part at the Jet Propulsion
Laboratory, under contract with the California Institute of
Technology (Caltech) funded by the National Aeronautics and Space
Administration (NASA) through the Sagan Fellowship Program executed
by the NASA Exoplanet Science Institute. MKL is supported by a CITA Postdoctoral Fellowship.
The authors are indebted to J. Carpenter, A. Isella, C. McNally, J. Oishi, and L. Ricci for thoughtful
suggestions, and to N. van der Marel for clarifying details
concerning the observations of Oph IRS 48. We thank also the anonymous referee for questions and comments that helped improve the work.

\end{document}